\documentclass[final,3p,times]{elsarticle}

\usepackage{amssymb}
\usepackage{MnSymbol}

\journal{Nuclear Physics B}

\begin{document}

\begin{frontmatter}



\title{Cross-correlations of coherent multiple Andreev reflections}


\author[label1]{Roman-Pascal Riwar}
\author[label2]{Driss M. Badiane}
\author[label1]{Manuel Houzet}
\author[label1]{Julia S. Meyer}
 \ead{julia.meyer@ujf-grenoble.fr}
\author[label3]{Yuli V. Nazarov}

\address[label1]{Univ. Grenoble Alpes, INAC-SPSMS, F-38000 Grenoble, France;\\
CEA, INAC-SPSMS, F-38000 Grenoble, France}
\address[label2]{Department of Physics, College of William and Mary, Williamsburg, Virginia 23187, USA}
\address[label3]{Kavli Institute of NanoScience, Delft University of Technology,
Lorentzweg 1, NL-2628 CJ, Delft, The Netherlands}

\begin{abstract}
We use the Landauer-B\"{u}ttiker scattering theory for electronic transport to calculate the current cross-correlations in a voltage-biased three-terminal junction with all superconducting leads. At low bias voltage, when charge transport is due to coherent multiple Andreev reflections, we find  large cross-correlations compared with their normal-state value. Furthermore, depending on the parameters that characterize the properties of the scattering region between the leads, the cross-correlations can reverse their sign with respect to the case of non-interacting fermionic systems.\\
{\tt Contribution for the special issue of Physica E in memory of Markus B\"{u}ttiker.}
\end{abstract}


\end{frontmatter}


\section{Introduction}
\label{Introduction}

Multiple Andreev reflections are the processes that explain how a dissipative charge transport can take place in a junction between superconducting leads at subgap voltages, $eV<2\Delta$, where $\Delta$ is the superconducting gap, and low temperatures~\cite{tinkham:1982}. Indeed, due to the energy gap in the excitation spectrum of conventional superconductors, the direct transfer of a quasiparticle between the leads is not possible in that voltage range. However, a subgap electron incident on a superconducting lead can be Andreev reflected as a hole, while a Cooper pair is created in the lead~\cite{andreev:1964}.  By performing $n$ successive Andreev reflections, it is  possible to transfer $\sim n/2$ Cooper pairs -- and one quasiparticle -- between two superconductors at voltage bias $eV>2\Delta/n$. This highly correlated process produces a rich subgap structure in the current-voltage characteristics, $I(V)$. Theoretically, this was predicted in incoherent~\cite{octavio:1983} and coherent~\cite{Averin1995,Bratus1995,Cuevas1996} ballistic junctions, as well as in diffusive junctions~\cite{bezuglyi:2001}. Experimentally, subgap structures in the $I(V)$-characteristics have been observed in a variety of systems, namely Josephson junctions based on exotic materials and nanoscale systems, such as semiconductor nanowires~\cite{xiang:2006,x2}, carbon nanotubes~\cite{buitelaar:2003,b2,b3}, and graphene flakes~\cite{heersche:2007}. Multiple Andreev reflections also result in a large shot noise, $S\sim q^* I$, both in the incoherent~\cite{bezuglyi1,nagaev,bezuglyi2} and coherent~\cite{noise-sns1,noise-sns2} regime. This effect can be ascribed to the divergence of the effective charge transferred in that process, $q^* =n e$ with $n\sim 2\Delta/(eV)$, as the voltage decreases.  The enhancement of the shot noise at low voltage bias was observed in tunnel~\cite{Diel:1997}, metallic~\cite{metalnoise,j2,j3}, and atomic point contact junctions~\cite{Cron:2001}. In the context of topological superconductivity, multiple Andreev reflections were recently discussed as a parity-changing process for the Majorana bound state that is formed in a topological Josephson junction~\cite{badiane1,badiane2}.

Further insight in the multiple Andreev reflection processes may be acquired through current cross-correlations in a multi-terminal geometry. As Markus B\"{u}ttiker demonstrated in his seminal paper on shot noise, the cross-correlations in non-interacting fermionic systems are always negative due to the Pauli principle~\cite{buttiker}. Later, several scenarios for a sign-reversal of the cross-correlations in the presence of interactions were proposed (see Ref.~\cite{buttiker2} for a review). For instance, the cross-correlations of the currents through two normal leads weakly contacted to a superconductor can be positive~\cite{datta,martin,torres,Samuelsson-buttiker,boerlin,Samuelsson-semiclassical,bignon,Melin,Feinberg}, due to a crossed Andreev reflection process in which an electron incident from one of the normal leads is Andreev-reflected to the other one~\cite{CAR,c2,c3,c4}. The possibility to use such cross-correlations for a signature of entanglement, due to the singlet-state of the crossed Andreev pair, was also discussed~\cite{lesovik}. Maximally positive cross-correlations -- meaning that they are exactly opposite to the autocorrelation -- in a topological superconductor in contact with two normal leads were predicted to be a signature of Majorana edge states~\cite{Nilsson2008,Law2009}, in the regime where the applied voltage exceeds the energy splitting between them. (By contrast, the cross-correlations vanish in the limit $V\to0$~\cite{BolechDemler}.) Recently, positive cross-correlations were measured in hybrid structures with tunnel junctions~\cite{Wei} and semiconducting nanowires~\cite{Das2012}. 

Motivated by these results, one of us studied the cross-correlations of multiple Andreev reflections in a normal chaotic dot attached to three superconducting leads~\cite{duhot:2009}. Depending on the coupling parameters, it was found that the cross-correlations acquire the same amplification factor  as the shot noise. Furthermore, a sign reversal at low voltage was predicted under certain conditions. However, this study was restricted to the incoherent regime, when the junction does not carry a supercurrent. This regime may occur when a small magnetic flux is applied to the junction to suppress the Josephson coupling between the leads. An incoherent regime is also expected at a temperature or an applied bias smaller than the gap, but larger than the energy scale that would characterize the induced minigap in the density of states of the dot in equilibrium. Cross-correlations in junctions with three superconducting terminals were measured recently~\cite{Kaviraj}. However, in that experiment only negative cross-correlations were observed.

The present work addresses the complementary coherent regime, where both an a.c.~Josephson effect and dissipative quasiparticle transport take place.
Phase-dependent multiple Andreev reflections  in the $I(V)$-characteristics  were investigated experimentally in a diffusive conductor~\cite{kutchinsky:1999} and  studied theoretically both in a single mode~\cite{Lantz2002} and a diffusive~\cite{Galaktionov2012} junction.  Our aim is to demonstrate that large positive cross-correlations may also occur in the coherent regime. The outline of the article is the following: in section \ref{scattering} we introduce the scattering theory of multiple Andreev reflections. Then we present the results for the current in section \ref{current} and for the noise and cross-correlations in section \ref{noise}. Section \ref{conclusion} contains the conclusions and outlook.

\section{Scattering theory of multiparticle Andreev reflection}
\label{scattering}

\begin{figure}
\begin{center}
\includegraphics[width=0.5\linewidth]{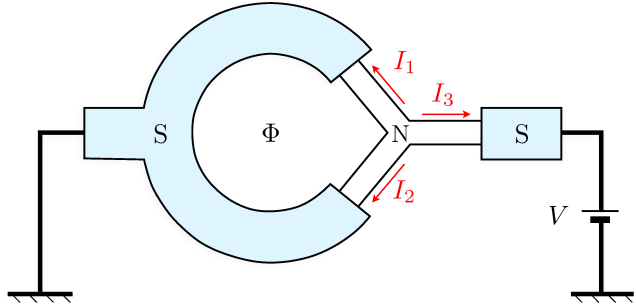}
\end{center}
\caption{Setup of a three-terminal superconducting junction. The superconducting leads are connected through a normal scattering region. A magnetic flux $\Phi$ applied through the loop formed between leads 1 and 2 allows controlling their superconducting phase difference, $\phi=2\pi \Phi/\Phi_0$, where $\Phi_0=hc/(2e)$ is the superconducting flux quantum. A voltage bias $V$ is applied to lead 3 while the leads 1 and 2 are grounded.}
\label{fig:setup}
\end{figure}

We consider a junction consisting of a normal scattering region connected to three superconducting leads, see Fig.~\ref{fig:setup}. Two leads (with labels $\alpha=1,2$) are grounded, while the third lead ($\alpha=3$) is biased with the voltage $V$. Furthermore, the superconducting loop geometry between leads 1 and 2 allows for imposing a superconducting phase difference $\phi$ that is tunable with the application of a magnetic flux through the loop. 

For simplicity, we assume that there is one channel per terminal, that the normal scattering region does not break time-reversal symmetry, and that there is no spin-orbit coupling in the system. Then the normal region can be characterized by a $3\times3$ scattering matrix, $\hat S(\varepsilon)=\{S_{\alpha\gamma}(\varepsilon)\}_{\alpha,\gamma=1,2,3}$. If it is shorter than the superconducting coherence length, the energy-dependence of $\hat S(\varepsilon)$ can be neglected. Assuming that leads 1 and 2 are symmetrically coupled to lead 3, the scattering matrix can be parametrized as
\begin{equation} 
\label{eq:scatt}
\hat S=\left(\begin{array}{ccc}
\sqrt{R}e^{ia}&\sqrt{D}&\sqrt{D_2}\\
\sqrt{D}&\sqrt{R}e^{ia}&\sqrt{D_2}\\
\sqrt{D_2}&\sqrt{D_2}&\sqrt{R_2}e^{ib}
\end{array}\right)\ ,
\end{equation}
up to irrelevant phases. Here $R=1-D-D_2$, $R_2=1-2D_2$, $a=\arccos[-D_2/(2\sqrt{R D})]$, and $b=\arccos[(D_2-2D)/(2\sqrt{R_2D})]$. The scattering matrix is thus parametrized by two real parameters: the transparency $D$ between leads 1 and 2 and  the transparency $2D_2$ between lead 3 and the other leads, with the constraints $D\leq1$ and $D_2\leq 2\sqrt{D}\,(1-\!\sqrt{D})$.

When $\phi=0$, the junction is equivalent to a two-terminal junction with transparency $2D_2$. However, the symmetry between leads 1 and 2  is broken at finite $\phi$. This can be related to the formation of a doubly-degenerate Andreev bound state with energy
\begin{equation}
E_A(\varphi_1,\varphi_2,\varphi_3)=\Delta\sqrt{1-D\sin^2\frac{\varphi_1-\varphi_2}2-D_2\left(\sin^2\frac{\varphi_2-\varphi_3}2+\sin^2\frac{\varphi_3-\varphi_1}2\right)}\ ,
\label{eq-EA} 
\end{equation}
when the junction is in equilibrium~\cite{Beenakker1991,Savinov2015}. Here $\varphi_\alpha$ is the superconducting phase of terminal $\alpha$ and $\phi=\varphi_1-\varphi_2$. For $D_2\ll1$ and finite voltage, a quasi-bound state with energy $E_{\rm qb}(\phi)=\Delta\sqrt{1-D\sin^2(\phi/2)}$ between leads 1 and 2 remains and affects the current as well as the noise and cross-correlations. The phase dependence of $I_3$, where $I_\alpha$ is the current going to contact $\alpha=1,2,3$, was studied in Ref.~\cite{Lantz2002}. Below we extend their results by computing the effect of multiple Andreev reflections on the current flowing between leads 1 and 2, $(I_1-I_2)/2$, as well as the current noise, $S_{33}$, and the cross-correlations, $S_{12}$.

To calculate the transport properties of the junction, we make use of the Landauer-B\"{u}ttiker theory, extended to describe hybrid junctions with superconducting leads~\cite{Averin1995,Bratus1995}.
For this, we first derive the wavefunctions associated with scattering states, which solve the time-dependent Bogoliubov-de Gennes equations describing the junction. In particular, the incoming and outgoing wavefunctions associated with an incoming electron-like state from lead $\beta$ at energy $E$ can be decomposed into their electron (e) and hole (h) amplitudes on the normal side of the interface between the junction and lead $\alpha$,
\begin{equation}
\hat \psi^{\rm in/out}_{{\rm e}\beta E}(t)=\left\{\psi^{\rm in/out,e}_{{\rm e}\beta E,1}(t),\psi^{\rm in/out,e}_{{\rm e}\beta E,2}(t),\psi^{\rm in/out,e}_{{\rm e}\beta E,3}(t),\psi^{\rm in/out,h}_{{\rm e}\beta E,1}(t),\psi^{\rm in/out,h}_{{\rm e}\beta E,2}(t),\psi^{\rm in/out,h}_{{\rm e}\beta E,3}(t)\right\}^T\ .
\end{equation}
Using their Fourier transform in energy space, $\hat \psi^{\rm in/out}_{{\rm e}\beta E}(t)=\int d\varepsilon/(2\pi)\, \hat \psi^{\rm in/out}_{{\rm e}\beta E}(\varepsilon)e^{-i\varepsilon t}$, we can relate the incoming and outgoing components of the electron part of the wave function through the scattering matrix $\hat S^{\rm e}$ for electrons. Namely,
\begin{equation}
\label{eq:WFARe}
\psi^{\rm out,e}_{{\rm e}\beta E,\alpha}(\varepsilon)=\sum_{\gamma=1,2,3} S^{\rm e}_{\alpha\gamma}
\psi^{\rm in,e}_{{\rm e}\beta E,\gamma}(\varepsilon+eV_\alpha-eV_\gamma)\ ,
\end{equation}
where $S^{\rm e}_{\alpha\gamma}=e^{i(\varphi_\alpha-\varphi_\gamma)/2}S_{\alpha\gamma}$, and $V_\alpha$ is the voltage at terminal $\alpha$. As specified above, $V_1=V_2=0$ and $V_3=V$. Note that we use units where $\hbar=1$.
 Similarly, we can relate the incoming and outgoing components of the hole part of the wave function,
\begin{equation}
\label{eq:WFARh}
\psi^{\rm out,h}_{{\rm e}\beta E,\alpha}(\varepsilon)=\sum_{\gamma=1,2,3} S_{\alpha\gamma}
^{\rm h}\psi^{\rm in,h}_{{\rm e}\beta E,\gamma}(\varepsilon-eV_\alpha+eV_\gamma)\ ,
\end{equation}
where $\hat S^{\rm h}=(\hat S^{\rm e})^*$ is the scattering matrix for holes.
Furthermore, the electron and hole components of the wavefunction at terminal $\alpha$ are related via Andreev reflections,
\begin{eqnarray}
\psi^{\rm in,e}_{{\rm e}\beta E,\alpha}(\varepsilon)&=&a(\varepsilon)
\psi^{\rm out,h}_{{\rm e}\beta E,\alpha}(\varepsilon)
+2\pi J(E) \delta_{\alpha\beta}\delta(\varepsilon-E)\ ,
\label{eq:WFSe}
\\
\psi^{\rm in,h}_{{\rm e}\beta E,\alpha}(\varepsilon)&=&a(\varepsilon)
\psi^{\rm out,e}_{{\rm e}\beta E,\alpha}(\varepsilon)\ .
\label{eq:WFSh}
\end{eqnarray}
Here the Andreev reflection amplitude is given as
\begin{equation}
a(\varepsilon)=\left\{\begin{array}{ll}
\varepsilon/\Delta-i\sqrt{1-\varepsilon^2/\Delta^2}\ , & |\varepsilon|<\Delta\ ,\\
\varepsilon/\Delta-{\rm sign}(\varepsilon)\sqrt{\varepsilon/\Delta^2-1}\ , & |\varepsilon|\geq\Delta\ .
\end{array}\right.
\end{equation}
The source term, $J(E)=\sqrt{1-|a(E)|^2}$, on the r.h.s.~of Eq.~\eqref{eq:WFSe} is a normalization coefficient, which ensures that the current carried by such a scattering state is $e/(2\pi)$.

Similarly, the incoming and outgoing wavefunctions, $\hat \psi^{\rm in/out}_{{\rm h}\beta E}(t)$, associated with an incoming hole-like state from lead $\beta$ at energy $E$ solve the same Equations \eqref{eq:WFARe}-\eqref{eq:WFSh}, except for the source term which appears in Eq.~\eqref{eq:WFSh} rather than Eq.~\eqref{eq:WFSe}. In particular, introducing the wavevector $\psi_{\nu,\alpha}=\left(\psi^{{\rm in, e}}_{\nu,\alpha},\psi^{{\rm out, h}}_{\nu,\alpha},\psi^{{\rm out, e}}_{\nu,\alpha},\psi^{{\rm in, h}}_{\nu,\alpha}\right)^T$, where $\nu=\{{\rm p},\beta, E\}$ for an incident particle of type ${\rm p}={\rm e/h}$, from lead $\beta$, and with energy $E$, one readily checks  the particle/hole symmetry relation
$\psi_{\bar\nu,\alpha}=\left(-\psi^{{\rm in, h}*}_{\nu,\alpha},\psi^{{\rm out, e}*}_{\nu,\alpha},-\psi^{{\rm out, h}*}_{\nu,\alpha},\psi^{{\rm in, e}*}_{\nu,\alpha}\right)^T$, 
where $\bar\nu=\{\bar {\rm p},\beta,-E\}$ with $\bar{\rm e}={\rm h}$ and $\bar{\rm h}={\rm e}$.

Using a Floquet decomposition, a solution of Eqs.~\eqref{eq:WFARe}-\eqref{eq:WFSh} may be written in the form
\begin{equation}
\label{eq:floquet}
\psi^{\rm in/out,{\rm p}}_{{\rm e}\beta E,\alpha}(\varepsilon)=2\pi \sum_{n=-\infty}^\infty\Psi^{\rm in/out,{\rm p}}_{{\rm e}\beta E,\alpha}(n)\delta(\varepsilon-E-neV)
\ .
\end{equation}
This reflects the periodic time-dependence of the Bogoliubov-de Gennes equations that describe the junction. Using the decomposition of Eq.~\eqref{eq:floquet}, we may write Eqs.~\eqref{eq:WFSe} and~\eqref{eq:WFSh} as
\begin{equation}
\Psi^{\rm in,{\rm p}}_{{\rm e}\beta E,\alpha}(n)=a_{n}(E) \Psi^{\rm out,\bar {\rm p}}_{{\rm e}\beta E,\alpha}(n)+J(E)\delta_{\alpha\beta}\delta_{n,0}\delta_{{\rm p},{\rm e}}\ ,
\end{equation}
with $a_{n}(E)=a(E+neV)$, and Eqs.~\eqref{eq:WFARe} and~\eqref{eq:WFARh} become
\begin{equation}
\Psi^{\rm out,{\rm e/h}}_{{\rm e}\beta E,\alpha}(n)=\sum_\gamma S_{\alpha\gamma}^{\rm e/h} 
\Psi^{\rm in,{\rm e/h}}_{{\rm e}\beta E,\gamma}(n\pm\delta_{\alpha,3}\mp\delta_{\gamma,3})\ .
\end{equation}

To obtain the current, noise, and cross-correlations, we need the operator for the current flowing to lead $\alpha$,
\begin{equation}
\label{eq:Iop}
I_\alpha(t)=ev_F {\cal C}^\dagger_\alpha(t)\sigma_z{\cal C}_\alpha(t)\ .
\end{equation}
Here $v_F$ is the Fermi velocity, 
\begin{equation}  
\sigma_z=\left(\begin{array}{cccc}
1&0&0&0\\
0&1&0&0\\
0&0&-1&0\\
0&0&0&-1
\end{array}\right)\ ,
\end{equation}
and ${\cal C}_\alpha$ is a Nambu spinor in particle-hole and in/out space, 
\begin{equation}
{\cal C}_\alpha(t)=\left(c_{{\rm in},\alpha\uparrow}(t),
c_{{\rm out},\alpha\downarrow}^\dagger(t),
c_{{\rm out},\alpha\uparrow}(t),
c_{{\rm in},\alpha\downarrow}^\dagger(t)\right)^T\ ,
\end{equation} 
where $c^\dagger_{{\rm in/out},\alpha\sigma}$ is a creation operator for an incoming/outgoing electron in lead $\alpha$ with spin $\sigma=\uparrow,\downarrow$. Using a Bogoliubov transformation, ${\cal C}_\alpha$ can be expressed in terms of the wavefunctions introduced above and the annihilation and creation operators $\gamma_{\nu,\sigma},\gamma^\dagger_{\nu,\sigma}$ for a Bogoliubov quasiparticle $\nu$ with spin $\sigma$, 
\begin{equation}
\label{eq:Bogo}
{\cal C}_\alpha(t)
=\sum_{\nu|E>0}
\psi_{\nu,\alpha}(t)
\gamma_{\nu,\uparrow}
+\psi_{\bar\nu,\alpha}(t)
\gamma_{\nu,\downarrow}^\dagger
\ .
\end{equation}
Defining
\begin{equation}
\gamma_\nu=\left\{\begin{array}{ll}
\gamma_{\nu ,\uparrow} \quad & {\rm if}\enspace E>0\ ,\\
\gamma_{\bar \nu,\downarrow}^\dagger & {\rm if}\enspace E<0\ ,\\
\end{array}\right.
\end{equation}
we may write Eq.~\eqref{eq:Iop} as
\begin{equation}
I_\alpha(t)=\frac e{2\pi}\sum_{\nu,\mu}M_{\nu\mu,\alpha}(t)\gamma^\dagger_\nu\gamma_\mu\ ,\qquad {\rm where} \qquad M_{\nu\mu,\alpha}(t)=\psi^\dagger_{\nu,\alpha}(t)\sigma_z\psi_{\mu,\alpha}(t)\ .
\label{eq-bogoliubov}
\end{equation}
Note that $\gamma_\nu$ obeys fermionic anticommutation relations, $\{\gamma_\nu,\gamma_\mu\}=0$ and $\{\gamma_\nu,\gamma^\dagger_\mu\}=\delta_{\nu\mu}$.
 
Assuming an equilibrium occupation of the scattering states, $\langle\gamma_\nu^\dagger\gamma_\nu\rangle=f(E)$, where $f(E)$ is the Fermi distribution function, we can compute the expectation value of the current, $\langle I_\alpha(t) \rangle$.  In particular, the d.c.~current reads
\begin{equation}
\label{eq:Idc}
\bar I_\alpha =I_\alpha^N-\frac e{2\pi}\int\limits_{-\infty}^\infty dE  \tanh\frac{E}{2T}
\left\{
2J(E)\Re\left[a(E)\Psi^{\rm out,h}_{{\rm e}\alpha E,\alpha}(0)\right]
+\sum_{\beta=1,2,3} \sum_{n=-\infty}^{\infty}
\left[\left(|a_{n}(E)|^2+1\right)\left( |\Psi^{\rm out,h}_{{\rm e}\beta E,\alpha} (n)|^2-|\Psi^{\rm out,e}_{{\rm e}\beta E,\alpha} (n)|^2\right)
\right]
\right\}
\ ,
\end{equation}
where 
\begin{equation}
I_\alpha^N=\frac {2e^2} h \sum_\beta |S_{\alpha\beta}|^2(V_\beta-V_\alpha)
\end{equation}
is the current flowing through the structure in the normal state. In particular, one finds $I_3^N=-I_1^N/2=-I_2^N/2=2D_2(2e^2/h)V$.

The current noise is given as
\begin{equation}
S_{\alpha\beta}(\omega)= \lim_{{\cal T}\to \infty}\frac1{2{\cal T}}\int_{-\cal T}^{\cal T} dt\int_{-\infty}^\infty d\tau\,e^{i\omega\tau}\langle \{\delta I_\alpha(t),\delta I_\beta(t+\tau)\}\rangle \ ,
\end{equation} 
where $\delta I_\alpha(t)= I_\alpha(t)-\langle I_\alpha(t)\rangle.$
Using Eq.~\eqref{eq-bogoliubov}, we find
\begin{equation}
\label{eq:S}
S_{\alpha\beta}(\omega)=\left(\frac{e}{2\pi}\right)^2 \lim_{{\cal T}\to \infty}\frac1{2{\cal T}}\int_{-\cal T}^{\cal T} dt\int_{-\infty}^\infty d\tau\,
\sum_{\nu\mu}M_{\nu\mu,\alpha}(t)M_{\mu\nu,\beta}(t+\tau)e^{i(\omega-E_\nu+E_\mu)\tau}\left[f_\nu(1-f_\mu)+f_\mu(1-f_\nu)\right]\ ,
\end{equation} 
where $E_\nu$ is the energy of state $\nu$ and $f_\nu=f(E_\nu)$.
By performing the time integrations in Eq.~\eqref{eq:S}, one may express  $S_{\alpha\beta}(\omega)$ explicitly in terms of the scattering wavefunctions, $\Psi^{\rm in/out,p}_{{\rm e}\gamma E,\delta}$, so that its numerical evaluation is straightforward. However, the expression is quite lengthy and we do not provide it in this article.

Note that, in the normal case and at zero temperature, one obtains the noise
\begin{equation}
S_{33}^N=\frac{4e^3}h 2D_2(1-2D_2)V,
\end{equation}
in accordance with the result for a two-terminal junction with transmission $2D_2$, and negative cross-correlations
\begin{equation}
S_{12}^N=-\frac{4e^3}h D_2^2V<0.
\label{eq-S_N}
\end{equation}
In the following sections, we compute the d.c.~current~\eqref{eq:Idc} and its correlators~\eqref{eq:S} for the superconducting three-terminal junction at zero temperature, associated with the scattering matrix~\eqref{eq:scatt}.

\section{Current}
\label{current}

\begin{figure}
\begin{center}
\includegraphics[width=0.85\linewidth]{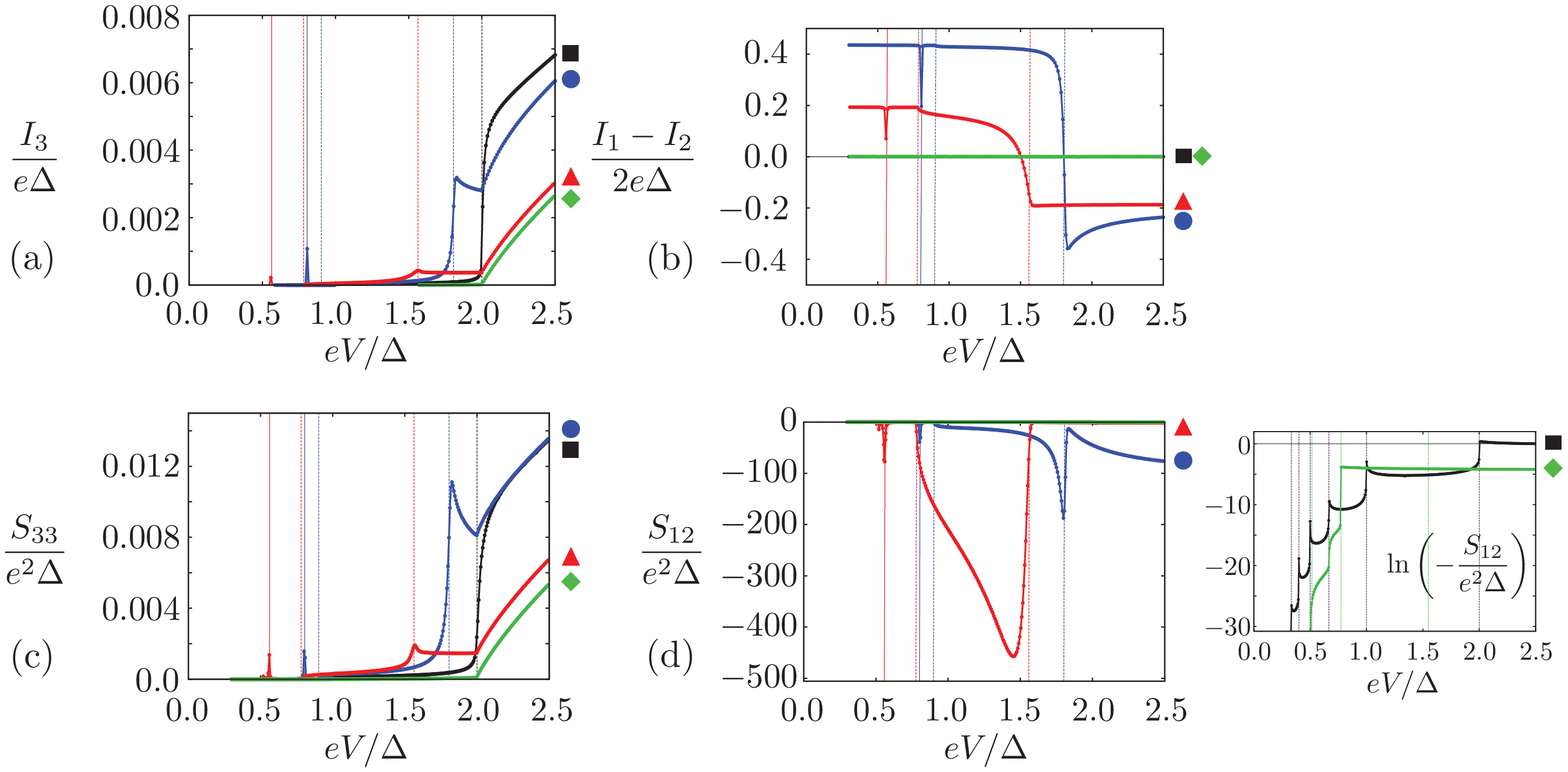}
\end{center}
\caption{Three-terminal junction characterized by the transparencies $D=0.7$ and $D_2=0.005$. We show (a) the current $I_3$, (b) the current $(I_1-I_2)/2$, (c) the noise $S_{33}$, and (d) the cross-correlations $S_{12}$ as a function of the bias voltage $V$ for different phases $\phi=0,0.5\pi,0.9\pi,\pi$ in black ($\filledmedsquare$), blue ($\medbullet$), red ($\blacktriangle$), and green ($\Diamondblack$), respectively.  For $\phi=0.5\pi, 0.9\pi$, the solid  vertical lines indicate the voltages $eV_{\rm Rabi}(\phi)$ while the dashed vertical lines inidcate the voltages $eV_{\rm DoS}(\phi)/n$ ($n=1,2$). In panels (a) and (c), in addition, $eV=2\Delta$ is shown. Note that the current $(I_1-I_2)/2$ in (b) is zero both for $\phi=0$ and $\phi=\pi$. In (d) we also show a zoom on the low-voltage regime to highlight the features due to multiple Andreev reflections, plotted on a logarithmic scale, for $\phi=0,\pi$. The vertical lines indicate the voltages $eV_n=2\Delta/n$ ($n\leq5$) for $\phi=0$ and $eV_{\rm DoS}(\phi)/n$ ($n\leq3$) for $\phi=\pi$.}
\label{fig-IS_D_0_7-I}
\end{figure}

\begin{figure}
\begin{center}
\includegraphics[width=0.85\linewidth]{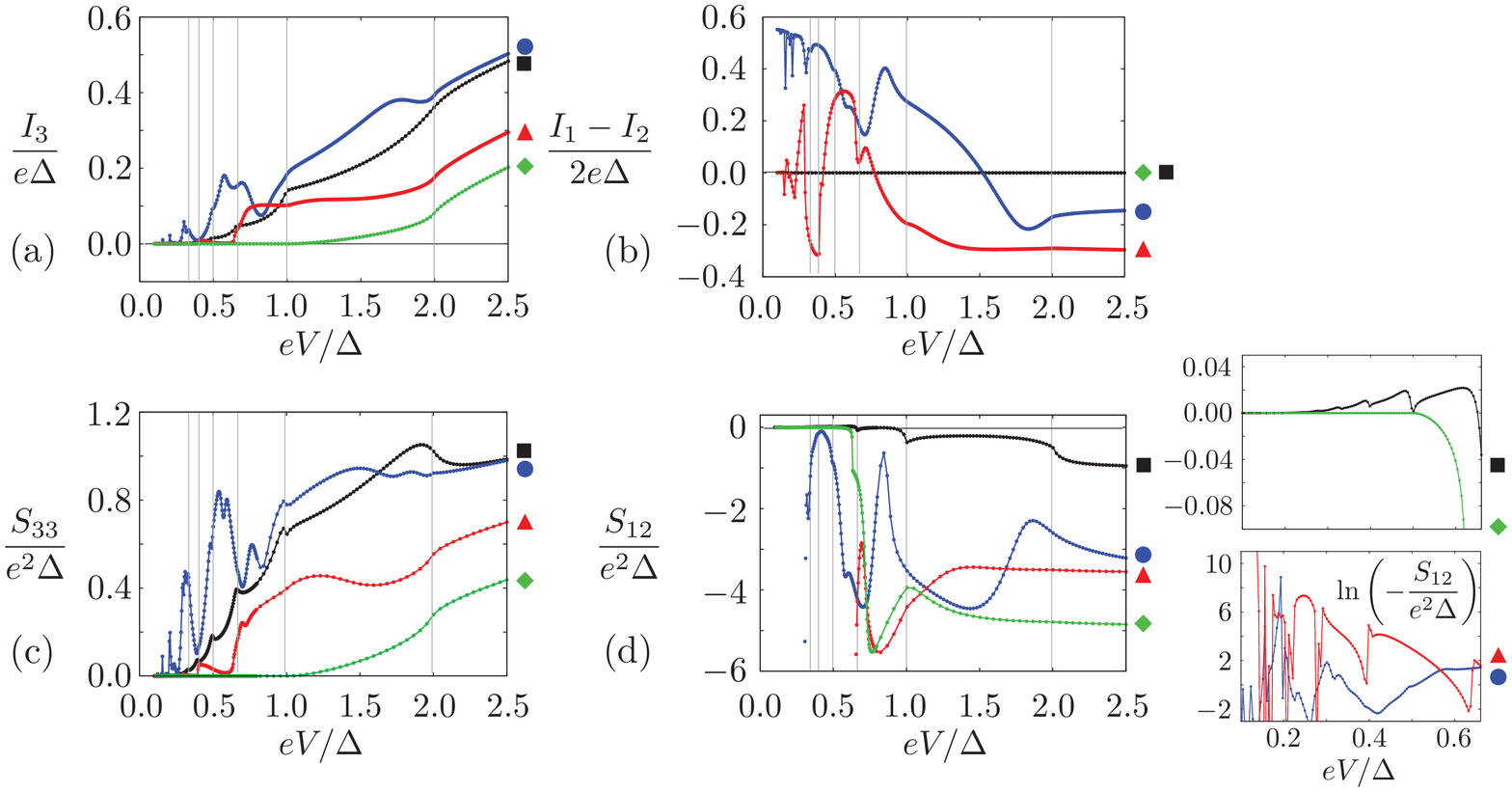}
\end{center}
\caption{Three-terminal junction characterized by the transparencies $D=0.7$ and $D_2=0.25$. We show (a) the current $I_3$, (b) the current $(I_1-I_2)/2$, (c) the noise $S_{33}$, and (d) the cross-correlations $S_{12}$ as a function of the bias voltage $V$ for different phases $\phi=0,0.5\pi,0.9\pi,\pi$ in black ($\filledmedsquare$), blue ($\medbullet$), red ($\blacktriangle$), and green ($\Diamondblack$), respectively.  The vertical lines indicate the voltages $eV_n=2\Delta/n$ for $n\leq6$. Note that the current $(I_1-I_2)/2$ in (b) is zero both for $\phi=0$ and $\phi=\pi$. In (d) we also show a zoom on the low-voltage regime to highlight the positive cross-correlations at $\phi=0$ (top) as well as the large negative cross-correlations, plotted on a logarithmic scale, for $\phi=0.5\pi,0.9\pi$ (bottom).}
\label{fig-IS_D_0_7-II}
\end{figure}

\begin{figure}
\begin{center}
\includegraphics[width=0.7\linewidth]{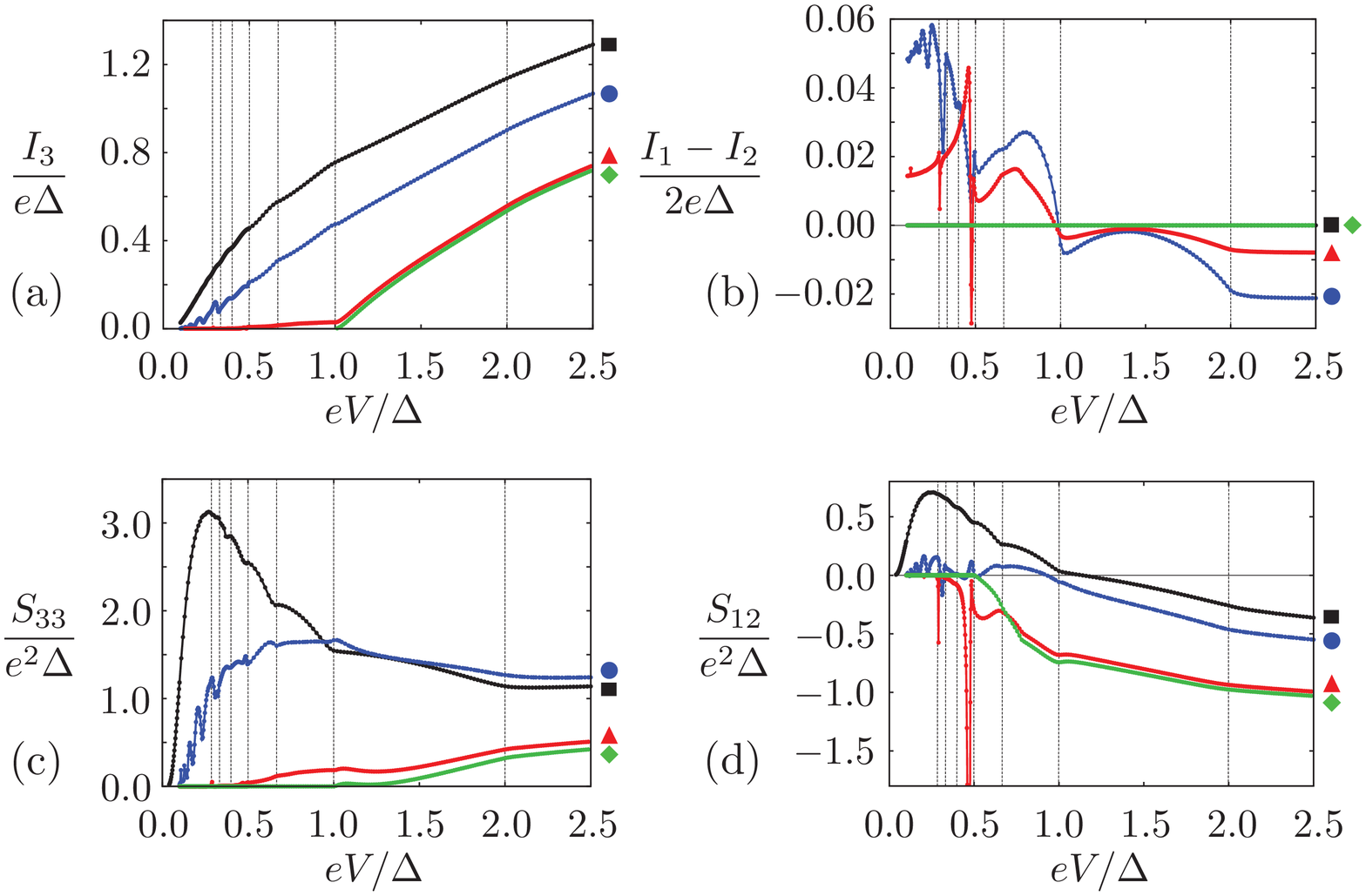}
\end{center}
\caption{Three-terminal junction characterized by the transparencies $D=0.12$ and $D_2=0.45$. We show (a) the current $I_3$, (b) the current $(I_1-I_2)/2$, (c) the noise $S_{33}$, and (d) the cross-correlations $S_{12}$ as a function of the bias voltage $V$ for different phases $\phi=0,0.5\pi,0.9\pi,\pi$  in black ($\filledmedsquare$), blue ($\medbullet$), red ($\blacktriangle$), and green ($\Diamondblack$), respectively. The vertical lines indicate the voltages $eV_n=2\Delta/n$ for $n\leq6$. Note that the current $(I_1-I_2)/2$ in (b) is zero both for $\phi=0$ and $\phi=\pi$.}
\label{fig-IS_D_0_12}
\end{figure}

Let us first discuss the current $I_3$ to the voltage biased lead 3. Panels (a) in Figs.~\ref{fig-IS_D_0_7-I}-\ref{fig-IS_D_0_12} show $I_3(V)$ for different values of the transmissions $D$ and $D_2$. The current displays multiple Andreev reflection features at voltages $eV_n=2\Delta/n$ as in the two-terminal case. 

In addition, in the tunneling regime at $D_2\ll1$ shown in Fig.~\ref{fig-IS_D_0_7-I}, it has resonances at the voltages $eV_{\rm Rabi} =E_{\rm qb}(\phi)$ and $eV_{\rm DoS} =\Delta+E_{\rm qb}(\phi)$, as discussed in Ref.~\cite{Lantz2002}. The two resonances have different origins. Namely, at the voltage $V_{\rm Rabi}$, the Josephson frequency $\omega_J=2eV_{\rm Rabi}$ matches the energy difference, $2E_{\rm qb}(\phi)$, between the situation where the quasi-bound state is occupied or empty. On the other hand, the voltage $V_{\rm DoS}$ corresponds to the situation where the gap edge of the lead 3 aligns with the energy of the empty quasi-bound state at energy $E_{\rm qb}(\phi)$. The values $E_{\rm qb}(\phi)$ for the phases $\phi$ shown in Fig.~\ref{fig-IS_D_0_7-I} are given in Table \ref{table}.

In Figs.~\ref{fig-IS_D_0_7-II} and \ref{fig-IS_D_0_12},  we show the results outside of the tunneling regime. As $D_2$ increases, the above features get washed out for most values of the phases $\phi$ as the variation of the Andreev bound state energy \eqref{eq-EA} with $\varphi_3$ increases. Namely, as a functions of $\varphi_3$, the bound state energy takes values in the interval $[E_A^-,E_A^+]$, 
where
\begin{equation}
E_A^\pm(\phi) =\Delta\sqrt{1-D\sin^2\frac\phi2-D_2\left(1\pm\left|\cos\frac\phi2\right|\right)}\ .
\end{equation}
Note that phases $\phi$ close to $\pi$ are special. Namely, at $\phi=\pi$, one finds $E_A^-(\pi)=E_A^+(\pi)$, and the adiabatic current between leads 1 and 2 is zero. Thus, for phases $\phi$ sufficiently close to $\pi$, one still finds a quasi-bound state at energy $\tilde E_{\rm qb}\approx\sqrt{1-D-D_2}$. The values $E_A^\pm$ for the transmissions $D$ and $D_2$ and phases $\phi$ chosen in Figs.~\ref{fig-IS_D_0_7-II} and~\ref{fig-IS_D_0_12} are given in Table \ref{table}. In Fig.~\ref{fig-IS_D_0_7-II}, additional structure away from the voltages $eV_n$ appears for values of $\phi$ such that the quasi-bound state carries an adiabatic current. However, it is difficult to give a clear interpretion in terms of the processes discussed above.  For Fig.~\ref{fig-IS_D_0_12}, the interval is too large to see any sharp features.

\begin{table}
\begin{center}
\renewcommand{\arraystretch}{1.5}
\begin{tabular}{|l||c|c|c|}
\hline
&Fig.~\ref{fig-IS_D_0_7-I}: $D=0.7,D_2=0.005$&Fig.~\ref{fig-IS_D_0_7-II}: $D=0.7, D_2=0.25$&Fig.~\ref{fig-IS_D_0_12}: $D=0.12,D_2=0.45$\\
\hline
\hline
$\phi=0$&$\frac{E_{\rm qb}}{\Delta}=1,\; \frac{I_{\rm ad}}{e\Delta}=0$&$\frac{E_A^-}{\Delta}\approx 0.71,\; \frac{E_A^+}{\Delta}=1,\; \frac{\bar I_{\rm ad}}{e\Delta}=0$&$\frac{E_A^-}{\Delta}\approx 0.32,\; \frac{E_A^+}{\Delta}=1,\; \frac{\bar I_{\rm ad}}{e\Delta}=0$\\
\hline
$\phi=0.5\pi$&$\frac{E_{\rm qb}}{\Delta}\approx0.81,\; \frac{I_{\rm ad}}{e\Delta}\approx0.43$&$\frac{E_A^-}{\Delta}\approx 0.47,\; \frac{E_A^+}{\Delta}\approx0.76,\; \frac{\bar I_{\rm ad}}{e\Delta}\approx0.56$&$\frac{E_A^-}{\Delta}\approx 0.41,\; \frac{E_A^+}{\Delta}\approx0.90,\; \frac{\bar I_{\rm ad}}{e\Delta}\approx0.05$\\
\hline
$\phi=0.9\pi$&$\frac{E_{\rm qb}}{\Delta}\approx0.56,\; \frac{I_{\rm ad}}{e\Delta}\approx0.19$&$\frac{E_A^-}{\Delta}\approx 0.17,\; \frac{E_A^+}{\Delta}\approx0.36,\; \frac{\bar I_{\rm ad}}{e\Delta}\approx0.37$&$\frac{E_A^-}{\Delta}\approx 0.60,\; \frac{E_A^+}{\Delta}\approx0.71,\; \frac{\bar I_{\rm ad}}{e\Delta}\approx0.01$\\
\hline
$\phi=\pi$&$\frac{E_{\rm qb}}{\Delta}\approx0.55,\; \frac{I_{\rm ad}}{e\Delta}=0$&$\frac{E_A^-}{\Delta}=\frac{E_A^+}{\Delta}\approx0.22,\; \frac{\bar I_{\rm ad}}{e\Delta}=0$&$\frac{E_A^-}{\Delta}=\frac{E_A^+}{\Delta}\approx0.66,\; \frac{\bar I_{\rm ad}}{e\Delta}=0$ \\
\hline
\end{tabular}
\caption{Energies of the quasi-bound state and adiabatic current for the values of transmissions and phases shown in Figs.~\ref{fig-IS_D_0_7-I}-\ref{fig-IS_D_0_12}.\label{table}}
\end{center}
\end{table}

\begin{figure}
\begin{center}
\includegraphics[width=0.8\linewidth]{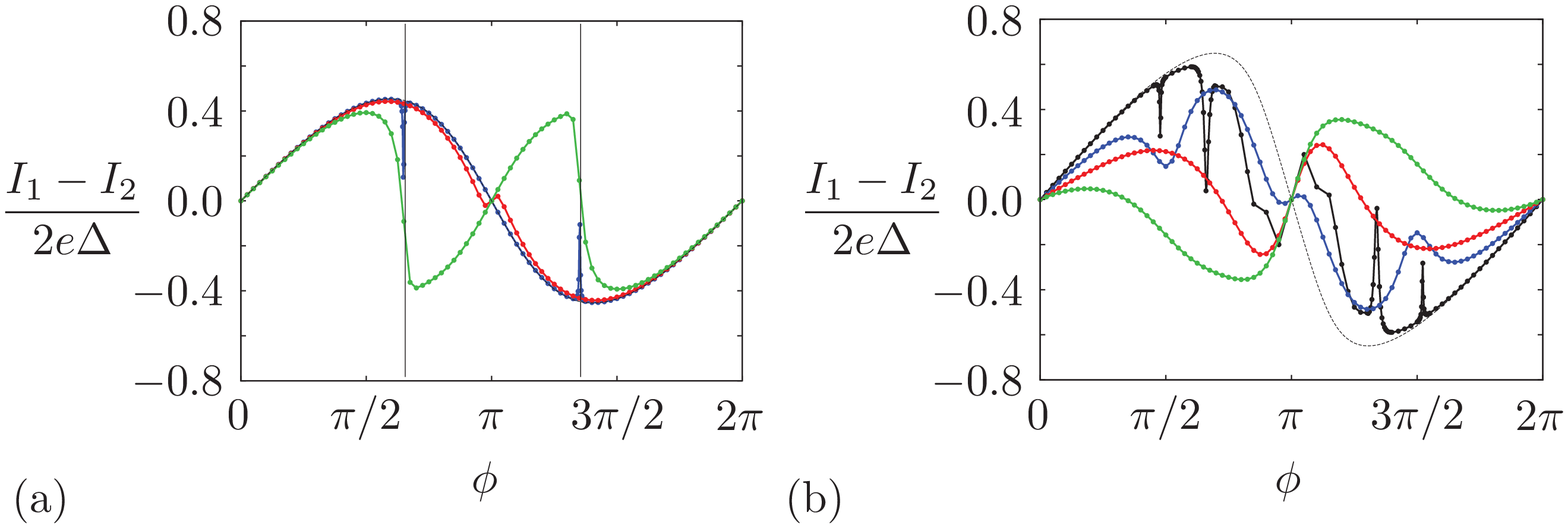}
\end{center}
\caption{Current $(I_1-I_2)/2$ as a function of the phase difference $\phi$ for different voltage biases, $eV/\Delta=0.16,0.7,1.16,1.7$ (black, blue, red, and green, respectively). The thin dashed line corresponds to the adiabatic current $\bar I_{\rm ad}$. The transparencies of the junction are given as $D=0.7$ whereas $D_2=0.005$ and $D_2=0.25$ in (a) and (b), respectively. The vertical lines in (a) indicate the phases $\phi_c^{\rm Rabi}(eV/\Delta=0.7)=\pi -\phi_c^{\rm DoS}(eV/\Delta=1.7)$ and  $2\pi-\phi_c^{\rm Rabi}(eV/\Delta=0.7)=\pi+\phi_c^{\rm DoS}(eV/\Delta=1.7)$. Note that the curve for $eV/\Delta=0.16$ in (a) is not displayed as it overlaps with the adiabatic current. }
\label{fig-I21}
\end{figure}

Next we turn to the current $(I_1-I_2)/2$, shown in panels (b) of Figs.~\ref{fig-IS_D_0_7-I}-\ref{fig-IS_D_0_12} as a function of $V$ and in Fig.~\ref{fig-I21} as a function of $\phi$. At small voltages and $D_2\ll1$, the current is given by the adiabatic current, 
\begin{equation}
I_{\rm ad}(\phi)=-2e\frac{\partial}{\partial\phi}E_{\rm qb}(\phi).
\end{equation}
As voltage is increased, two different features are observed. These features are linked with the resonances in the current $I_3$ discussed above. At $eV_{\rm Rabi}  =E_{\rm qb}(\phi)$, one finds a suppression of the current, corresponding to an average occupation of the doubly-degenerate quasi-bound state with one quasiparticle. At $eV >eV_{\rm DoS}=\Delta+E_{\rm qb}(\phi)$, the injection of quasiparticles from lead 3 leads to a reversal of the current, corresponding to a double occupation of the quasi-bound state. Due to the phase dependence of $E_{\rm qb}(\phi)$, the reversal appears first around the phase $\pi$ and then spreads over the entire phase interval as voltage is further increased.

In summary, there are several different voltage regimes: at $eV<E_{\rm qb}(\pi)=\Delta\sqrt{1-D}$ the current is adiabatic. At voltages $E_{\rm qb}(\pi) <eV<\Delta$, one observes resonances at $\phi_c^{\rm Rabi}$ and $2\pi-\phi_c^{\rm Rabi}$ with $\phi_c^{\rm Rabi}=2\arcsin\sqrt{[1-(eV/\Delta)^2]/D}$. At voltages $\Delta<eV<\Delta+E_{\rm qb}(\pi)$, the current is again adiabatic.  At voltages $\Delta+E_{\rm qb}(\pi)<eV<2\Delta$, the current is reversed in the phase interval $\phi\in[\pi-\phi_c^{\rm DoS},\pi+\phi_c^{\rm DoS}]$ with $\phi_c^{\rm DoS}=\pi-2\arcsin\sqrt{[1-(eV/\Delta-1)^2]/D}$. Finally, at voltages $eV>2\Delta$, the current is reversed for all phases.

As can be seen in Fig.~\ref{fig-I21}(a), for $D_2\ll1$, the current  $(I_1-I_2)/2$  is well decribed by the above considerations, except for the sign reversal in a narrow interval around $\phi=\pi$ observed in the regime $\Delta<eV<\Delta+E_{\rm qb}(\pi)$. This can be traced to the fact that ionization processes, where a quasiparticle escapes the bound state, are very weak for phases close to $\pi$ because the dependence of the bound state energy on the applied voltage vanishes at $\phi=\pi$. As a consequence, higher order processes with a threshold at $E_{\rm DoS}(\phi)/n$ are sufficient to establish a significant population of the quasi-bound state. The sign reversal seen here occurs at $eV=E_{\rm DoS}(\phi)/2$.

As discussed above, the above features get washed out for most values of the phases $\phi$ when $D_2$ increases. The adiabatic current in that case is given as
\begin{equation}
\bar I_{\rm ad}(\phi)=-2e\frac{\partial}{\partial\phi}\bar E_A(\phi),
\end{equation}
where $\bar E_A(\phi)=\int_0^{2\pi}d\varphi_3 \,E_A(\varphi_1,\varphi_2,\varphi_3)/(2\pi)$. The current outside of the tunneling regime is shown in Fig.~\ref{fig-I21}(b). Here the current  $(I_1-I_2)/2$ follows the adiabatic approximation only for small voltages and phases close to zero. Facilitated by multiple Andreev reflections, a sign reversal for phases close to $\pi$ is observed at all voltages. Furthermore, we see resonances at voltages smaller than $\Delta$. As the energy of the bound state $E_A$ strongly varies with $\varphi_3$, it is not obvious why the resonances at low voltage are so narrow.

In Figs.~\ref{fig-IS_D_0_7-I}-\ref{fig-IS_D_0_12}, panels (b), we note that for voltages $eV>2\Delta$ the current $(I_1-I_2)/2$ flattens out at a value that is opposite in sign as compared to the adiabatic current and smaller in magnitude. 

\section{Noise and cross-correlations }
\label{noise}

We now turn to the noise and cross-correlations. The noise is shown in Figs.~\ref{fig-IS_D_0_7-I}-\ref{fig-IS_D_0_12}, panels (c). When $D_2$ is large and multiple Andreev reflections are important, the noise is strongly enhanced at low voltages, see Fig.~\ref{fig-IS_D_0_12}(c).

As shown in Fig.~\ref{fig-IS_D_0_12}(d), the cross-correlations at $\phi=0$ may be large and positive at low voltages. In particular, this is the case when $D_2$ is large. It is due to crossed Andreev reflections, where Cooper pairs are split between lead 1 and 2. At large voltages, when superconducting correlations are less important, the cross-correlations are always negative. The large-voltage asymptotic approaches the normal state result, Eq.~\eqref{eq-S_N}.

In addition to this main result, we notice that the noise and cross-correlations display additional features that are related with the features discussed above for the current. Namely, the multiple Andreev reflection features at voltages $eV_n=2\Delta/n$ are visible in the noise and cross-correlations as well. While their position is independent of $\phi$, their visibility varies. In particular, they become more and more pronounced as $\phi$ increases from $0$ to $\pi$. In the tunneling regime, additional features due to the presence of a quasi-bound state are present, see Fig.~\ref{fig-IS_D_0_7-I}(d). The Rabi oscillations occurring near voltages $eV_{\rm Rabi}$ lead to large negative cross-correlations. This feature is similar to the large supercurrent noise that was predicted in equilibrium Josephson junctions due to the thermal fluctuation in the Andreev bound state occupation~\cite{noise-eq-sns1,noise-eq-sns2}. Large negative cross-correlations are also observed in the interval $[V_{\rm DoS}(\phi)/2,V_{\rm DoS}(\phi)]$, where the average occupation of the quasi-bound state change from $0$ to a value close to $2$. (The associated change of sign of $I_1-I_2$ can be seen in panel (b) of Fig.~\ref{fig-IS_D_0_7-I}.) When $D_2$ increases, the negative cross-correlations at finite phase difference become less pronounced as seen in Fig.~\ref{fig-IS_D_0_12}(d). Note that, even at large voltages, a dependence of the noise and cross-correlations on the phase $\phi$ remains in the form of a voltage-independent excess contribution.


\section{Conclusion}
\label{conclusion}

Three-terminal Josephson junctions realize positive cross-correlations at low voltages. These correlations are strongly enhanced compared to the case where only one of the leads is superconducting due to the process of multiple Andreev reflections. Here we studied the coherent regime where supercurrents and dissipative quasiparticle currents lead to an interesting interplay that manifests itself in the current as well as in the noise and cross-correlations.

As a next step, it would be interesting to address the crossover between coherent and incoherent multiple Andreev reflection regimes in multichannel hybrid junctions. The methods introduced in our work could also be helpful to test recent predictions related with topological aspects of the Andreev subgap spectrum in multiterminal Josephson junctions~\cite{VanHeck,Riwar}.

We dedicate this article to the memory of Markus B\"uttiker. His groundbreaking contributions to mesoscopic physics have shaped the field and remain an inspiration for our research. 

This work was supported by ANR,
through grants ANR-11-JS04-003-01 and ANR-12-BS04-0016-03, and by the Nanosciences Foundation in Grenoble, in the frame of its Chair of Excellence program.




\begin{thebibliography}{00}


\bibitem{tinkham:1982} 
	T. M. Klapwijk, G. E. Blonder, and M. Tinkham, 
	Physica B \& C {\bf 109-110}, 1657 (1982).
	
\bibitem{andreev:1964}
	A.F. Andreev,
	Zh. Eksp. Teor. Fiz. {\bf 46}, 1823 (1964)
	[Sov. Phys. JETP {\bf 19}, 1228 (1964)].

\bibitem{octavio:1983} 
	M. Octavio, M. Tinkham, G. E. Blonder, and T. M. Klapwijk, 
	Phys. Rev. B {\bf 27}, 6739 (1983).

\bibitem{Averin1995} 
	D. Averin and A. Bardas, 
	Phys. Rev. Lett. {\bf 75}, 1831 (1995).

\bibitem{Bratus1995} 
	E. N. Bratus, V. S. Shumeiko, and G. Wendin, 
	Phys. Rev. Lett. {\bf 74}, 2110 (1995).

\bibitem{Cuevas1996} 
	J. C. Cuevas, A. Mart\'in-Rodero, and A. L. Yeyati, 
	Phys. Rev. B {\bf 54}, 7366 (1996).

\bibitem{bezuglyi:2001}
	E. V. Bezuglyi, E. N. Bratus, V. S. Shumeiko, G. Wendin, and H. Takayanagi, 
	Phys. Rev. B {\bf 62}, 14439 (2000).
	
	
\bibitem{xiang:2006}
	Y.~J. Doh, J. A. van Dam, A. L. Roest, E. P. A. M. Bakkers, L. P. Kouwenhoven, and S. De Franceschi,
	Science {\bf 309}, 272 (2005).
	
\bibitem{x2}	
	Jie Xiang, A. Vidan, M. Tinkham, R. M. Westervelt, and C. M. Lieber,
	Nature Nanotechnology {\bf 1}, 208 (2006).

\bibitem{buitelaar:2003}
	M. R. Buitelaar, W. Belzig, T. Nussbaumer, B. Babic, C. Bruder, and C. Sch\"onenberger.
	Phys. Rev. Lett. {\bf 91}, 057005 (2003).
	
\bibitem{b2}	
	P. Jarillo-Herrero, J. A. van Dam, and L. Kouwenhoven, Nature {\bf 439}, 953 (2006).
	
\bibitem{b3}
	H. I. J\o rgensen, K. Grove-Rasmussen, T. Novotn\'y, K. Flensberg, and P. E. Lindelof, 
	Phys. Rev. Lett. {\bf 96}, 207003 (2006).

\bibitem{heersche:2007}
	H. B. Heersche, P. Jarillo-Herrero, J. B. Oostinga, L. M. K. Vandersypen, and A. F. Morpurgo,
	Nature {\bf 446}, 56 (2007).

\bibitem{bezuglyi1}
	E. V. Bezuglyi, E. N. Bratus', V. S. Shumeiko, and G. Wendin, 
	Phys. Rev. Lett. {\bf 83}, 2050 (1999).
	
\bibitem{nagaev} 
	K. E. Nagaev, 
	Phys. Rev. Lett. {\bf 86}, 3112 (2001).
	
\bibitem{bezuglyi2}
	E. V. Bezuglyi, E. N. Bratus, V. S. Shumeiko, and G. Wendin, 
	Phys. Rev. B {\bf 63}, 100501(R) (2001).	
	
\bibitem{noise-sns1} 
	Y. Naveh and D. V. Averin, 
	Phys. Rev. Lett. {\bf 82}, 4090 (1999).	
	
\bibitem{noise-sns2} 
	J. C. Cuevas, A. Martin-Rodero, and A. Levy Yeyati, 
	Phys. Rev. Lett. {\bf 82}, 4086 (1999).
		
\bibitem{Diel:1997}
	P. Dieleman, H. G. Bukkems, T. M. Klapwijk, M. Schicke, and K. H. Gundlach, 
	Phys. Rev. Lett. {\bf 79}, 3486 (1997).

\bibitem{metalnoise} 
	X. Jehl, P. Payet-Burin, C. Baraduc, R. Calemczuk, and M. Sanquer, 
	Phys. Rev. Lett. {\bf 83}, 1660 (1999).
	
\bibitem{j2} 	
	T. Hoss, C. Strunk, T. Nussbaumer, R. Huber, U. Staufer, and C. Sch\"onenberger, 
	Phys. Rev. B {\bf 62}, 4079 (2000).
	
\bibitem{j3} 		
	C. Hoffmann, F. Lefloch, M. Sanquer, and B. Pannetier, 
	Phys. Rev. B {\bf 70}, 180503 (2004).

\bibitem{Cron:2001}
	R. Cron, M. F. Goffman, D. Esteve, C. Urbina, 
	Phys. Rev. Lett. {\bf 86}, 4104 (2001).
	
\bibitem{badiane1}
	D. M. Badiane, M. Houzet, and J. S. Meyer,
	Phys. Rev. Lett. {\bf 107}, 177002 (2011).
	
\bibitem{badiane2}
	D. M. Badiane, L. I. Glazman,
M. Houzet, and J. S. Meyer, C.R. Physique {\bf 14}, 840 (2013).
	
\bibitem{buttiker}
	M. B\"uttiker, 
	Phys. Rev. B {\bf 46}, 12 485 (1992).
	
\bibitem{buttiker2}
	M. B\"uttiker, 
	{\it in} "Quantum Noise", edited by Yu. V. Nazarov and Ya. M. Blanter (Kluwer, 2002),
	arXiv:cond-mat/0209031.
	
\bibitem{datta}
	M. P. Anantram and S. Datta, 
	Phys. Rev. B {\bf 53}, 16 390 (1996).

\bibitem{martin} 
	T. Martin, 
	Phys. Lett. A {\bf 220}, 137 (1996).
	
\bibitem{torres}
	J. Torr\`{e}s and T. Martin, 
	Eur. Phys. J. B {\bf 12}, 319 (1999).

\bibitem{Samuelsson-buttiker} 
	P. Samuelsson and M. B\"uttiker,
	Phys. Rev. Lett. {\bf 89}, 046601 (2002).

\bibitem{boerlin}
	J. B\"orlin, W. Belzig, and C. Bruder
	Phys. Rev. Lett. {\bf 88}, 197001 (2002).
	
\bibitem{Samuelsson-semiclassical} 
	P. Samuelsson and M. B\"uttiker,
	Phys. Rev. B. {\bf 66}, 201306(R) (2002).
%
\bibitem{bignon}
	G. Bignon, M. Houzet, F. Pistolesi, and F. W. J. Hekking, 
	Europhys. Lett. {\bf 67}, 110 (2004).	
	
\bibitem{Melin}	
R. M\'elin, C. Benjamin, and T. Martin,
Phys. Rev. B {\bf 77}, 094512 (2008).

\bibitem{Feinberg}
M. Fl\"oser, D. Feinberg, and R. M\'elin,
Phys. Rev. B {\bf 88}, 094517 (2013).

\bibitem{CAR}
	J. M. Byers and M. E. Flatt\'{e}, 
	Phys. Rev. Lett. {\bf 74}, 306 (1995).
	
\bibitem{c2}	
	G. Deutscher and D. Feinberg, 
	Appl. Phys. Lett. {\bf 76}, 487 (2000).
	
\bibitem{c3}		
	P. Recher, E. V. Sukhoroukov, and D. Loss, 
	Phys. Rev. B {\bf  63} 165314 (2001).
	
\bibitem{c4}		
	G. Falci, D. Feinberg, and F. W. J. Hekking, 
	Europhys. Lett. {\bf 54}, 255 (2001).	

\bibitem{lesovik}
	N. M. Chtchelkatchev, G. Blatter, G. B. Lesovik, and T. Martin,
	Phys. Rev. B {\bf 66}, 161320 (2002).

\bibitem{Nilsson2008}
	J. Nilsson, A. R. Akhmerov, and C. W. J. Beenakker,
	Phys. Rev. Lett. {\bf 101}, 120403 (2008).
	
\bibitem{Law2009}
	K. T. Law, P. A. Lee, and T. K. Ng,
	Phys. Rev. Lett. {\bf 103}, 237001 (2009).
	
\bibitem{BolechDemler}
C. J. Bolech and E. Demler, Phys. Rev. Lett. {\bf 98}, 237002 (2007).

\bibitem{Wei}
	J. Wei and V. Chandrasekhar, 
	Nature Phys. {\bf 6}, 494 (2010).

\bibitem{Das2012}
	A. Das, Y. Ronen, M. Heiblum, D. Mahalu, A. V. Kretinin, and H. Shtrikman, 
	Nature Commun. {\bf 3}, 1165 (2012).
	
\bibitem{duhot:2009}
	S. Duhot, F. Lefloch, and M. Houzet,
	Phys. Rev. Lett. {\bf 102}, 086804 (2009).

\bibitem{Kaviraj}
	B. Kaviraj, O. Coupiac, H. Courtois, and F. Lefloch, 
	Phys. Rev. Lett {\bf 107}, 077005 (2011).

\bibitem{kutchinsky:1999}
	J. Kutchinsky, R. Taboryski, C. B. S\"orensen, J. B. Hansen, and P. E. Lindelof, 
	Phys. Rev. B {\bf 56}, R2932 (1997).

\bibitem{Lantz2002}
	J. Lantz, V. S. Shumeiko, E. Bratus, and G. Wendin, 
	Phys. Rev. B {\bf 65}, 134523 (2002).
	
\bibitem{Galaktionov2012}
	A. V. Galaktionov, A. D. Zaikin, and L. S. Kuzmin, 
	Phys. Rev. B {\bf  85}, 224523 (2012).
	
\bibitem{Beenakker1991}
	C. W. J. Beenakker, 
	Phys. Rev. Lett. {\bf 67}, 3836 (1991).

\bibitem{Savinov2015}	
	D. A. Savinov,
	Physica C {\bf 509}, 22 (2015).
	
\bibitem{noise-eq-sns1} 
	D. Averin and H. T. Imam, 
	Phys. Rev. Lett. {\bf 76}, 3814 (1996).

\bibitem{noise-eq-sns2} 
	A. Mart\'{i}n-Rodero, A. Levy Yeyati, and F. J. Garc\'{i}a-Vidal, 
	Phys. Rev. B {\bf 53}, 8891(R) (1996).
	
\bibitem{VanHeck}
	B. van Heck, S. Mi, and A. R. Akhmerov, 
	Phys. Rev. B {\bf 90}, 155450 (2014).

\bibitem{Riwar}
	R.-P. Riwar, M. Houzet, J. S. Meyer, and Yu. V. Nazarov,
	arXiv:1503.06862.

\end{thebibliography}


\end{document}